\documentclass[aps,prb,twocolumn,superscriptaddress,showpacs]{revtex4-1}
\usepackage{graphicx}
\usepackage{dcolumn}
\usepackage{bm}
\newcommand{\Vec}[1]{\mbox{\boldmath$#1$}}

\begin{document}


\title{Understanding the re-entrant superconducting 
phase diagram of an iron-pnictide 
Ca$_4$Al$_2$O$_6$Fe$_2$(As$_{1-x}$P$_x$)$_2$ }


\author{Hidetomo Usui}
\affiliation{Department of Physics,
Osaka University, 1-1 Machikaneyama, Toyonaka, Osaka 560-0043, Japan}

\author{Katsuhiro Suzuki}
\affiliation{Department of Engineering Science,
The University of Electro-Communication, Chofu, Tokyo 182-8585, Japan}

\author{Kazuhiko Kuroki}
\affiliation{Department of Physics,
Osaka University, 1-1 Machikaneyama, Toyonaka, Osaka 560-0043, Japan}

\author{Nao Takeshita}
\affiliation{National Institute of Advanced Industrial Science 
and Technology (AIST), Central-2, 1-1-1, Umezono, Tsukuba, Ibaraki 305-8568, 
Japan}

\author{Parasharam Maruti Shirage }
\affiliation{National Institute of Advanced Industrial Science 
and Technology (AIST), Central-2, 1-1-1, Umezono, Tsukuba, Ibaraki 305-8568, 
Japan}

\author{Hiroshi Eisaki}
\affiliation{National Institute of Advanced Industrial Science 
and Technology (AIST), Central-2, 1-1-1, Umezono, Tsukuba, Ibaraki 305-8568, 
Japan}

\author{Akira Iyo}
\affiliation{National Institute of Advanced Industrial Science 
and Technology (AIST), Central-2, 1-1-1, Umezono, Tsukuba, Ibaraki 305-8568, 
Japan}

\date{\today}

\begin{abstract}
Recently, a very rich phase diagram has been obtained for an 
iron-based superconductor Ca$_4$Al$_2$O$_6$Fe$_2$(As$_{1-x}$P$_x$)$_2$.
It has been revealed that nodeless $(x\sim 0)$ and nodal $(x=1)$ 
superconductivity are separated by an antiferromagnetic phase.
 Here we study the origin of this peculiar phase diagram 
using a five orbital 
model constructed from first principles band calculation, 
and applying the fluctuation 
exchange approximation  assuming spin fluctuation 
mediated pairing.
At $x=1$, there are three hole Fermi surfaces, but the 
most inner one around the wave vector $(0,0)$ has strong $d_{X^2-Y^2}$ 
orbital character, unlike in LaFeAsO, where the most inner Fermi surface 
has $d_{XZ/YZ}$ character. Since the Fermi surfaces around $(0,0)$, $(\pi,0)$
and $(\pi,\pi)$ all have $d_{X^2-Y^2}$ orbital character, the 
repulsive pairing interaction mediated by the spin fluctuations gives rise to 
a frustration in momentum space, thereby degrading superconductivity 
despite the bond angle being close to the regular tetrahedron angle.
As $x$ decreases and the bond angle is reduced, the inner hole Fermi surface 
disappears, but the frustration effect still remains because the 
top of the band with $d_{X^2-Y^2}$ character lies close to the 
Fermi level. On the other hand, the loss of the Fermi surface itself 
gives rise to a very good nesting of the Fermi surface because the 
number of electron and hole Fermi surfaces are now the same. 
The pairing interaction frustration and the good nesting combined 
favors antiferromagnetism over superconductivity.
Finally for $x$ close to 0, the band sinks far below the Fermi level, 
reducing the frustration effect, so that superconductivity is enhanced. 
There, the Fermi surface nesting is 
also lost to some extent, once again favoring superconductivity 
over antiferromagnetism. In order to see whether the present 
theoretical scenario is consistent with the actual nature of the 
competition between superconductivity and antiferromagnetism, 
we also perform  hydrostatic pressure experiment for 
Ca$_4$Al$_2$O$_6$Fe$_2$(As$_{1-x}$P$_x$)$_2$.
In the intermediate $x$ regime where 
antiferromagnetism occurs at ambient pressure, 
applying hydrostatic pressure smears out the antiferromagnetic transition, 
but superconductivity does not take place. This supports our scenario that 
superconductivity is suppressed by the momentum space frustration in the 
intermediate $x$ regime, 
apart from the presence of the antiferromangnetism.
\end{abstract}


\maketitle

\section{Introduction}
The discovery of the iron-based superconductors has given great impact 
not only because of the high $T_c$, but also because it raises 
a fundamental question on the pairing mechanism 
in a class of high $T_c$ materials other than the 
cuprates\cite{Kamihara2008,Ren2008}. 
In fact, a spin fluctuation mediated pairing mechanism was 
proposed right after the discovery of superconductivity\cite{Mazin2008,2008PRL}.
One interesting and important feature 
of the iron-based superconductors is the 
relationship between the superconducting transition temperature and the 
lattice structure, in particular, the Fe-Pn (Pn:Pnictogen) 
positional relationship\cite{Lee2008,Mizuguchi2010}.  
Lee ${\it et \  al.}$ have 
experimentally shown that $T_c$ systematically varies with the Fe-Pn-Fe bond 
angle, and takes its maximum around 109$^\circ$, 
at which the pnictogen atoms form 
a regular tetrahedron\cite{Lee2008}. 
On the other hand the strength of the low lying 
spin fluctuation seems to be stronger for materials with 
bond angle smaller than the regular tetrahedron angle, i.e., those 
materials with moderate to low $T_c$.
\cite{Kinouchi2011,Ishidarev,Nakai,Fujiwara,Mukuda}.

Ca$_4$Al$_2$O$_6$Fe$_2$As$_2$ is particularly interesting 
in this context. This material was synthesized by Shirage {\it et al.}
\cite{Shirage2010} as a variation of a series of materials 
that have thick perovskite layers in between FeAs layers\cite{Ogino}.
This material is particularly interesting from the lattice structure viewpoint 
in that it has a very small Fe-As-Fe bond angle of 102$^\circ$.
It has been revealed by NMR experiment\cite{Kinouchi2011} in the 
normal state that the spin fluctuation 
is very strong in this material despite the moderate $T_c$ of about 28K.
The $1/T_1$ measurement in the superconducting state suggests that 
the gap is fully open with a sign change between electron and hole 
Fermi surfaces, namely, an fully gapped $s\pm$ state\cite{Kinouchi2011}.

Theoretically, we have previously explained this correlation 
among the lattice structure, the spin fluctuations, 
and the superconducting $T_c$/gap structure within 
the spin-fluctuation-mediated pairing scenario using a five orbital model 
obtained for the hypothetical lattice structure of LaFeAsO
\cite{Kuroki2009a,Usui2011}. 
We have concluded that superconductivity 
is strongly affected by the Fermi surface multiplicity, and 
the spin fluctuation 
is strongly affected by the hole Fermi surface around the wave vector 
$(\pi,\pi)$ in the unfolded Brillouin zone.
It has been found that the number of Fermi surface is 
controlled by the Fe-Pn-Fe 
bond angle or the pnictogen height. When the bond angle is large 
(low pnictogen height), two hole Fermi surfaces around the wave vector $(0,0)$ 
are present. In this case, a low $T_c$ nodal $s$-wave paring 
or $d$-wave paring takes place\cite{Graser,Kuroki2009a,DHLee,Thomale}. 
As the bond angle $\alpha$ decreases, the hole Fermi surface 
appears around $(\pi,\pi)$, and we now have three hole Fermi surfaces. 
This is what has been noticed as an effect of increasing the pnictogen height
\cite{Singh,Vildosola,Kuroki2009a}. 
the interaction between the electron and the hole Fermi 
surfaces gives rise to a high 
$T_c$ $s\pm$-wave paring, where the gap is fully open but changes sign between 
electron and hole Fermi surfaces as was first proposed in 
ref. \onlinecite{Mazin2008}. 
Upon reducing $\alpha$ even further 
(and thus increasing the high pnictogen height), 
the inner hole Fermi surface around $(0,0)$ 
disappears, and again there are only two hole Fermi surfaces. 
Here, the good Fermi surface nesting gives rise to a strong spin fluctuation, 
while the superconducting $T_c$ of the gapped $s\pm$ state 
remains to be moderate  because of the reduction of the 
scattering processes.
Thus, superconductivity is 
optimized in the intermediate bond angle regime around 110$^\circ$,  
where the Fermi surface multiplicity is maximized. 

However, there are some experimental observations that 
seem to be beyond the understanding of the above mentioned theory.
For example, the phosphide version of this 42622 material, 
Ca$_4$Al$_2$O$_6$Fe$_2$P$_2$, has a lower  
$T_c$ of 17K\cite{Shirage2010} although the bond angle is nearly 109$^\circ$,
which is very close to the regular tetrahedron bond angle.
In the phosphides, the Fe-Pn bond length is generally reduced compared to the 
arsenides, so the density of states tends to be smaller.   
Therefore, the phosphides and the arsenides 
do not have to obey the same $T_c$ vs. bond angle dependence.
Still, there seems to be some effect that suppresses $T_c$ in 
Ca$_4$Al$_2$O$_6$Fe$_2$P$_2$, 
considering the fact that (i) $T_c=17K$ is nearly the same as 
Sr$_4$Sc$_2$O$_6$Fe$_2$P$_2$ with a much larger bond angle\cite{Ogino}, 
(ii) the band structure calculation for 
Ca$_4$Al$_2$O$_6$Fe$_2$P$_2$ by Kosugi {\it et al.} \cite{Kosugi} shows that 
the number of hole Fermi surfaces is three, i.e., the 
inner Fermi surface is not lost as opposed to Ca$_4$Al$_2$O$_6$Fe$_2$As$_2$, 
and the Fermi surface multiplicity is maximized,  
(iii) an NMR experiment for Ca$_4$Al$_2$O$_6$Fe$_2$P$_2$ 
suggests presence of nodes in the superconducting gap (or a very small gap 
at some portions of the Fermi surface)\cite{Kinouchi2012}.

Quite recently, an interesting observation has been made for 
Ca$_4$Al$_2$O$_6$Fe$_2$(As$_{1-x}$P$_x$)$_2$, an isovalent 
doping material, where As is (partially) replaced by P. 
As mentioned above, the end materials at $x=0$ and $x=1$ are 
both superconductors. In between these two phases, 
antiferromagnetism takes place in the intermediate regime of the P 
content $x$, and separates the two superconductivity phases of 
$x \sim 0$\cite{Shirage2010} and  $x \sim 1$.\cite{Shirage,Kinouchi2012} 
Therefore, Ca$_4$Al$_2$O$_6$Fe$_2$(As$_{1-x}$P$_x$)$_2$  
varies from a fully gapped superconducting state to an antiferromagnetism 
and finally to a nodal superconducting state.

In this paper, we study this peculiar behavior of 
superconductivity and antiferromagnetism in 
Ca$_4$Al$_2$O$_6$Fe$_2$(As$_{1-x}$P$_x$)$_2$ from 
a lattice structure and band structure point of view.  
We calculate the band structure of the hypothetical lattice 
structure of Ca$_4$Al$_2$O$_6$Fe$_2$As$_2$ and construct an effective 
five band model exploiting the maximally localized Wannier orbitals.
In varying the bond angle in a wide range 
while fixing the bond length, we find that the most inner hole Fermi surface 
around the wave vector $(0,0)$ in the unfolded Brillouin zone changes its 
orbital character from $XZ/YZ$ to  $X^2-Y^2$ just before it 
disappears. Then the Fermi surfaces around the wave vectors 
$(0,0)$, $(\pi,0)$ and $(\pi,\pi)$ will all (partially) have 
$X^2-Y^2$ orbital character. This is the Fermi surface configuration 
for Ca$_4$Al$_2$O$_6$Fe$_2$P$_2$.
Since the spin fluctuation mediated 
pairing interaction tends to change the sign of the superconducting 
gap between portions of the Fermi surface having similar orbital 
character, this will give rise to a frustration in momentum space,
degrading superconductivity. 
 We also explain the competition between 
superconductivity and antiferromagnetism in 
Ca$_4$Al$_2$O$_6$Fe$_2$(As$_{1-x}$P$_x$)$_2$. Around the intermediate region of 
$x$, the most inner Fermi surface is lost, but the top of the band 
still lies close to the Fermi level. In this situation, the momentum 
space frustration effect is still strong, and the superconductivity is 
suppressed. At the same time and independently, 
the Fermi surface itself is nearly 
perfectly nested since there are now two electron and two hole Fermi surfaces 
with the same total area (for zero doping). For smaller $x$, the 
band that gives rise to the frustration sinks far below the 
Fermi level, and superconductivity again takes over antiferromagnetism.

In order to see whether the present theoretical scenario 
is consistent with the actual 
nature of the superconductivity-antiferromagnetism 
competition in the present material, 
we also perform hydrostatic pressure experiment.
In the intermediate $x$ regime, superconductivity 
does not take place under pressure although 
the  pressure smears out the antiferromagnetic transition.
This experiment further supports our view that 
in the intermediate $x$ regime, superconductivity is 
suppressed by some origin other than the antiferromagnetism 
itself, which in our view is the momentum space frustration.

\section{Band structure}
\subsection{Original lattice structure}
\begin{figure}
\includegraphics[width=8.5cm]{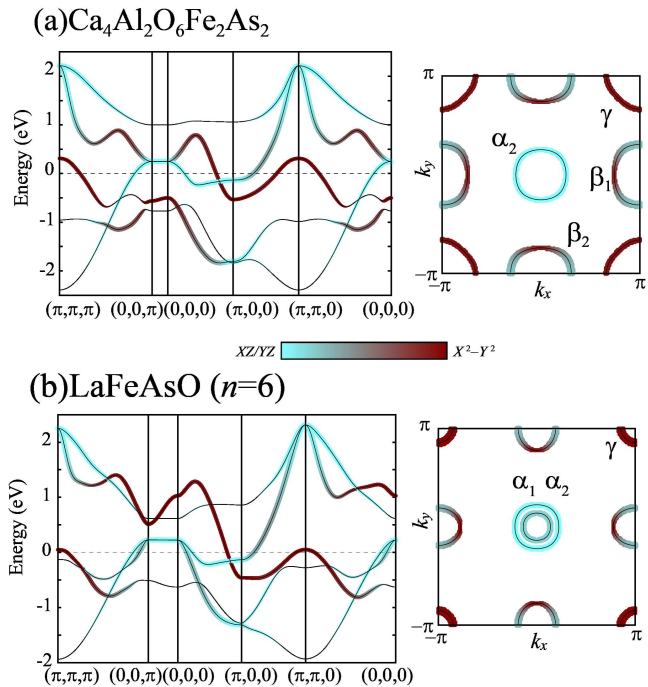}
\caption{(a)The band structure and the Fermi surface of (a) LaFeAsO and 
(b) Ca$_4$Al$_2$O$_6$Fe$_2$As$_2$. 
The thickness of the lines represents the weight of 
the  $X^2-Y^2$ or $XZ/YZ$ orbital characters.
\label{fig:1}
}
\end{figure}
We first calculate the band structure of 
Ca$_4$Al$_2$O$_6$Fe$_2$As$_2$, which was first performed in 
ref.\onlinecite{Miyake2010}, and compare it to that of LaFeAsO. 
We adopt the lattice structure determined experimentally\cite{Shirage2010},
where the Fe-As-Fe bond angle $\alpha$ is 102$^{\circ}$ and the 
pnictogen height $h_{\rm Pn}$ measured from the iron plane is 1.5\AA.
The first principles band calculation is performed 
using the Quantum-Espresso package\cite{pwscf}, 
and we construct a five orbital tight-binding 
Hamiltonian\cite{2008PRL}  
exploiting the maximally localized Wannier functions\cite{MaxLoc}. 
The five Wannier orbitals consist mainly of Fe $3d$ and As $4p$ 
orbitals, and 
these orbitals have five different symmetries ($d_{XY}$, $d_{YZ}$,
$d_{ZX}$, $d_{3Z^2-R^2}$ and $d_{X^2-Y^2}$), 
where $X,Y$ refer to the direction of 
rotated by 45 degrees from the Fe-Fe direction $x,y$. 
The multi orbital tight binding 
Hamiltonian is expressed as 
\begin{eqnarray}
H_0 = \sum_{\sigma}\sum_{i,\mu} \varepsilon_{\mu}c^{\dagger}_{i\mu\sigma}c_{i\mu\sigma} + 
\sum_{\sigma}\sum_{ij,\mu\nu}t^{\mu\nu}_{ij}c^{\dagger}_{i\mu\sigma}c_{j\nu\sigma},
\end{eqnarray}
where $t^{\mu\nu}_{ij}$ is the hopping, 
$i,j$ denote the sites and $\mu,\nu$ specify the orbitals.
We define the band filling $n$ as the number of electrons per site, 
where $n=6$ refers to the non-doped case. The Fermi surfaces shown 
in Fig.\ref{fig:1} are those for the $k_z=0$ plane and $n=6$.

As pointed out in ref.\onlinecite{Miyake2010}, a large difference between the 
band structure of the two materials is the number of hole Fermi surfaces.
In LaFeAsO, there are two hole Fermi surfaces around 
the wave vector $(k_x,k_y)=(0,0)$ originating from 
the $XZ/YZ$ orbitals, and one hole Fermi surface around $(\pi,\pi)$ 
originating from the $X^2-Y^2$  orbital. 
In Ca$_4$Al$_2$O$_6$Fe$_2$As$_2$ by contrast,  
one of the hole Fermi surfaces around (0,0) ($\alpha_1$) is missing.
This difference is due to the position of the upper 
portion of the $X^2-Y^2$ band along 
$(0,0,0)-(0,0,\pi)$ indicated by the short arrows in Fig.\ref{fig:1}.
This band lies above the $XZ/YZ$ bands in LaFeAsO, while it lies below the 
$XZ/YZ$ bands in Ca$_4$Al$_2$O$_6$Fe$_2$As$_2$. We will come back to this 
point in more detail in the next subsection.
Another large difference between the two materials is the 
strength of the two dimensionality.
The band structure of Ca$_4$Al$_2$O$_6$Fe$_2$As$_2$ has strong 
two-dimensionality due to the large block layer, namely, the dispersion 
of the upper $X^2-Y^2$ band 
along $(0,0,0)-(0,0,\pi)$ is much smaller than in LaFeAsO. 

\begin{figure}
\includegraphics[width=8.5cm]{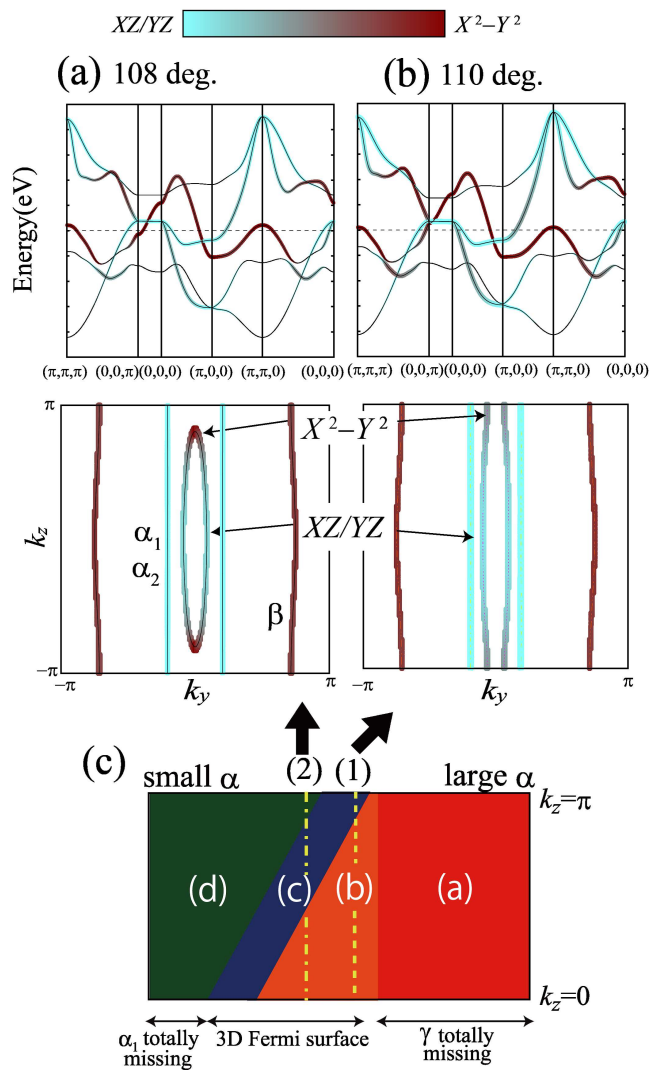}
\caption{The band structure (top) and the vertical cut of the Fermi surface of 
LaFeAsO for hypothetical lattice structures with (a)$\alpha=108^\circ$ or 
(b) $\alpha=110^\circ$. 
The thickness of the lines represents the weight of 
the  $X^2-Y^2$ or $XZ/YZ$ orbital characters.
(c) A schematic figure representing the band 
structure/Fermi surface configuration in the $\alpha$-$k_z$ plane.
(a)-(d) correspond to the configurations shown in Fig.\ref{fig:3}.\label{fig:2}
}
\end{figure}

\begin{figure}
\includegraphics[width=8.5cm]{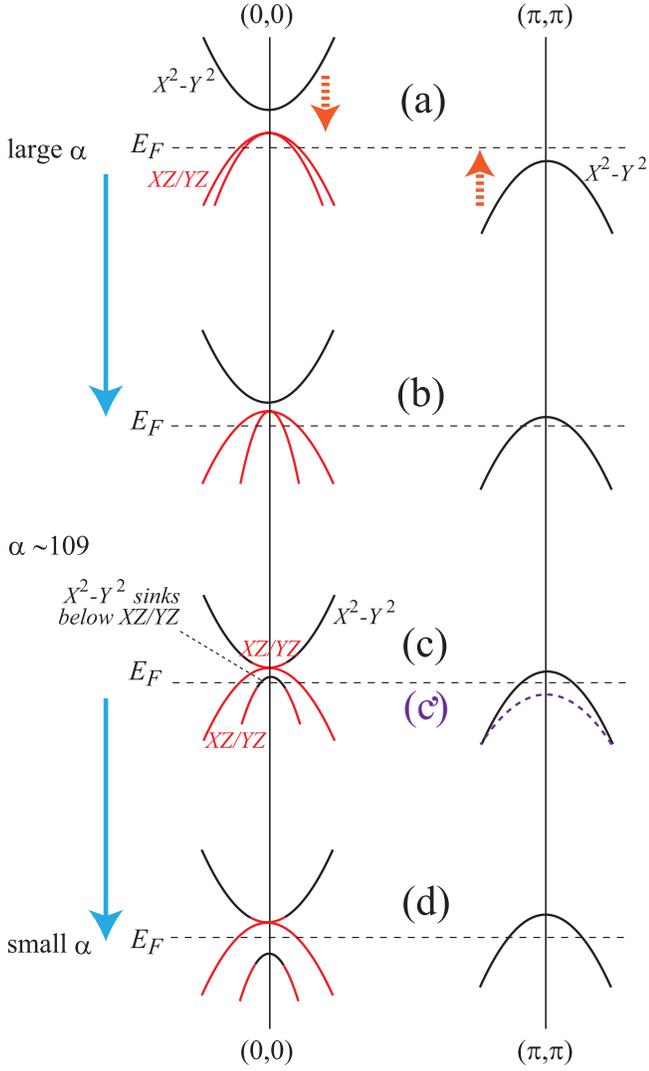}
\caption{A schematic figure of the band structure variation against 
the bond angle $\alpha$. 
The solid black (solid red) portions indicate the bands with strong $X^2-Y^2$ ($XZ/YZ$) orbital character.\label{fig:3}}
\end{figure}

\subsection{Bond angle variation}

As was done in refs.\onlinecite{Miyake2010,Usui2011}, 
we discuss the bond angle dependence of the band structure.
Before going into Ca$_4$Al$_2$O$_6$Fe$_2$As$_2$, we summarize the 
bond angle variation of the band structure of LaFeAsO, which was 
discussed in detail in refs.\onlinecite{Usui2011,SUST}.
In Fig.\ref{fig:2}, we show the band structure of LaFeAsO for the 
hypothetical lattice structures with smaller bond angles than in the 
original lattice structure with 113$^\circ$. 
The lower portion of the $X^2-Y^2$ band 
around $(\pi,\pi)$ rises up upon reducing the bond angle, and at the 
same time the upper $X^2-Y^2$ band along 
$(0,0,0)-(0,0,\pi)$ comes down and partially sinks below the $XZ/YZ$ bands 
at 108$^\circ$.
This variation of the bands is schematically summarized in Fig.\ref{fig:3}
\cite{SUST}. When the upper $X^2-Y^2$ band sinks below the $XZ/YZ$ bands,
reconstruction of the band structure takes place, and one of the 
hole Fermi surface is lost for sufficiently small bond angle
(configuration (d)). It is important to note that just before the 
$\alpha_1$ hole Fermi surface is lost, 
the $X^2-Y^2$ orbital character strongly mixes into 
the $\alpha_1$ Fermi surface (configuration (c)). 
Due to the three dimensional dispersion of the upper $X^2-Y^2$ band in 
LaFeAsO, the disappearance of the $\alpha_1$ Fermi surface is $k_z$ dependent,
so that the Fermi surface becomes three dimensional for a certain 
bond angle regime, as shown in the left panel 
of Fig.\ref{fig:2}(c). Even when the 
Fermi surface itself is two dimensional, the orbital character can 
change along the $k_z$ direction as shown in the right panel.
This $k_z$ dependence of the Fermi surface configuration of LaFeAsO 
is schematically summarized in Fig.\ref{fig:2}(c)\cite{SUST}.

\begin{figure}
\includegraphics[width=8.5cm]{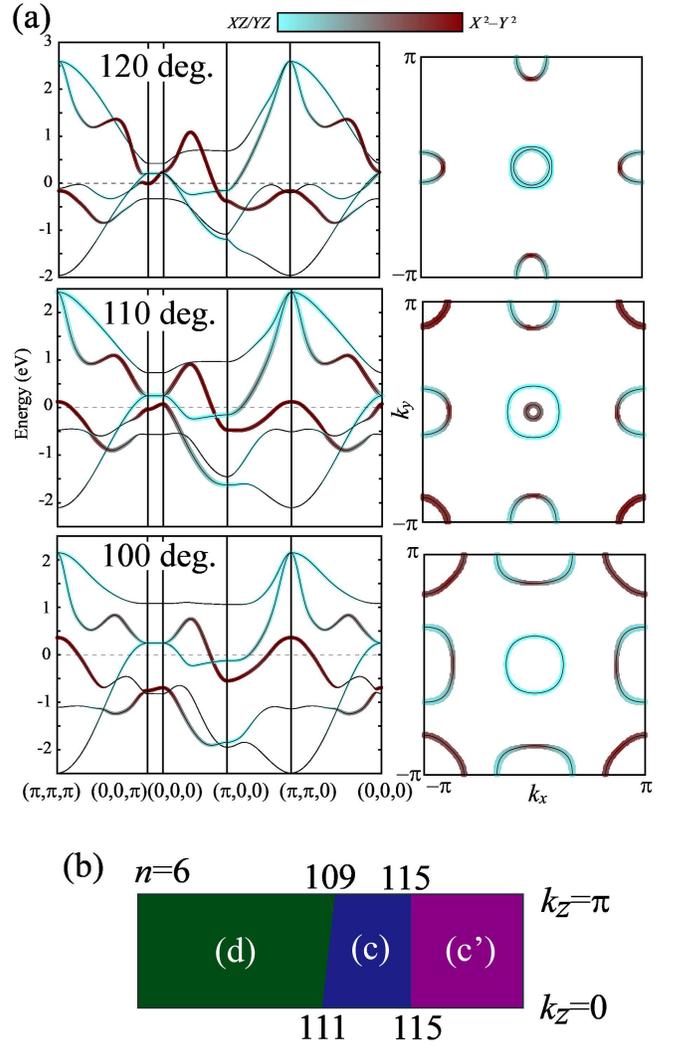}
\caption{(a)The band structure (left) and the Fermi surface at $k_z=0$ 
(right) of Ca$_4$Al$_2$O$_6$Fe$_2$As$_2$ for hypothetical lattice 
structures with $\alpha=120^\circ$, $110^\circ$, and $100^\circ$.
The thickness of the lines represents the weight of 
the  $X^2-Y^2$ or $XZ/YZ$ orbital characters.\label{fig:4}}
\end{figure}

Baring this in mind, we now move on to 
Ca$_4$Al$_2$O$_6$Fe$_2$As$_2$. Although this was analyzed in detail 
in ref.\onlinecite{Miyake2010}, here we put more focus on the orbital character 
of the most inner hole Fermi surface.
In Fig.\ref{fig:4}, we show the band structure variation upon 
decreasing the bond angle from 120$^\circ$ to 100$^\circ$. 
(The original lattice structure is 102$^\circ$.) 
It is interesting to note that 
most portion of the upper $X^2-Y^2$ band sinks below the 
$XZ/YZ$ bands even at 120$^\circ$. Therefore, 
a Fermi surface configuration that does not occur in 
LaFeAsO takes place. This is schematically shown in 
Fig.\ref{fig:3} as configuration (c$'$). 
As the bond angle is reduced, the $\gamma$ Fermi surface 
around $(\pi,\pi)$ appears, followed by the disappearance of the $\alpha_1$ 
hole Fermi surface. This disappearance occurs in a narrow bond angle regime 
between 111$^\circ$ to 109$^\circ$ due to the strong two dimensionality.
The Fermi surface configuration variation for 
Ca$_4$Al$_2$O$_6$Fe$_2$As$_2$ is summarized in Fig.\ref{fig:4}(b).
Here it is important to note that configuration (b) in Fig.\ref{fig:3} 
does not appear in this case, namely, the inner $\alpha_1$ 
hole Fermi surface always has some mixture of $X^2-Y^2$ orbital 
component in the regime where three hole Fermi surfaces exist.
As we shall see, 
this will affect the conclusion in our previous paper\cite{Usui2011},  
i.e., superconductivity is optimized in the bond angle 
regime in which the multiplicity of the hole Fermi surface is maximized.

\begin{figure}
\includegraphics[width=8.5cm]{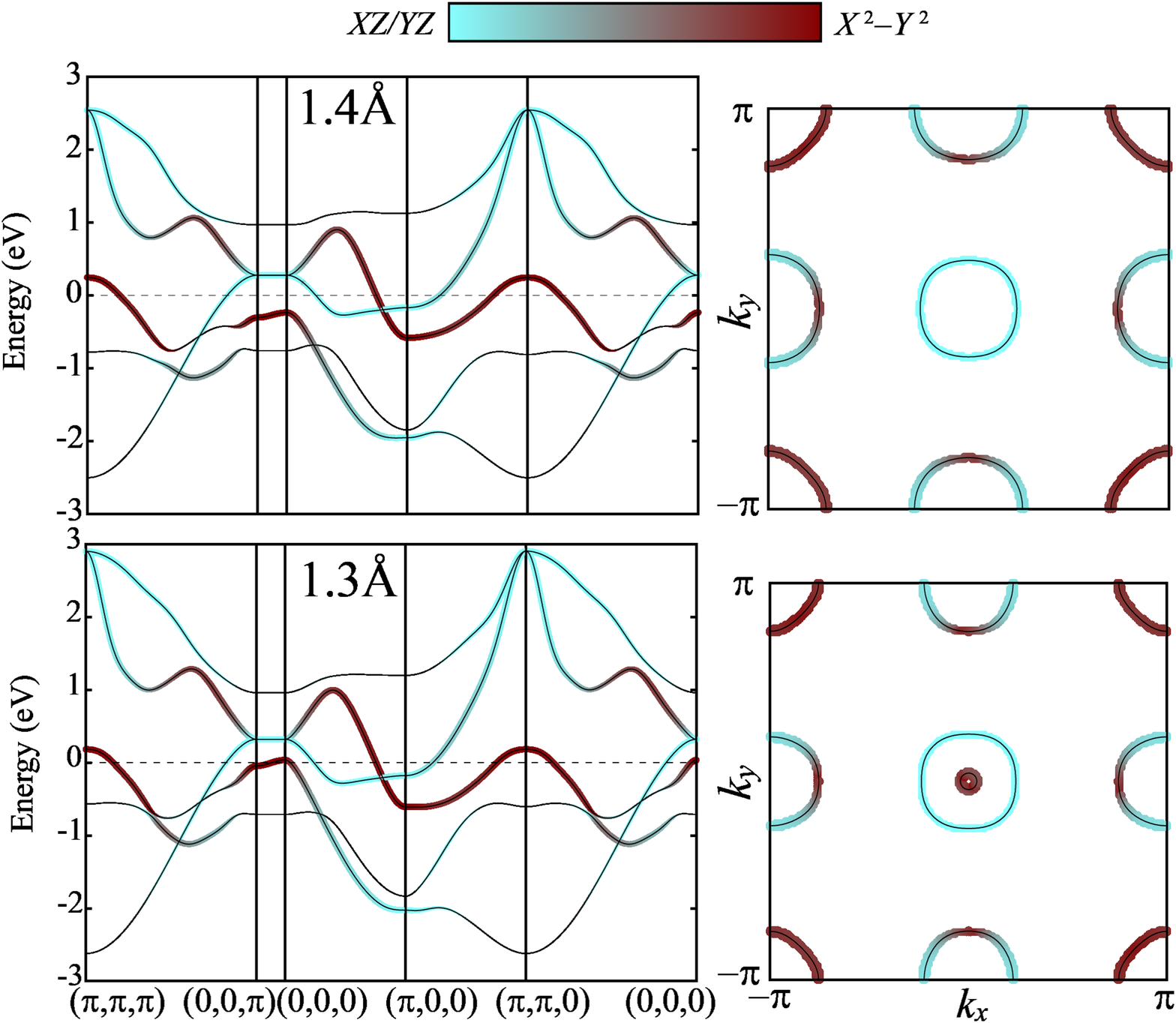}
\caption{The band structure and the Fermi surface of hypothetical lattice 
structure in Ca$_4$Al$_2$O$_6$Fe$_2$As$_2$ with varying pnictogen height 
while fixing lattice parameter $a$.\label{fig:5}}
\end{figure}

\subsection{Height variation}
Upon partially replacing As by P in 
Ca$_4$Al$_2$O$_6$Fe$_2$(As$_{1-x}$P$_x$)$_2$, 
the bond angle reduction is accompanied by the 
increase in the Fe-Pn bond length. 
Therefore, the lattice parameter $a$ hardly 
decreases, while the pnictogen height measured from the 
iron plane largely decreases  
from 1.5 to 1.3\AA \  as the P content $x$ increases from 0 to 1.
We show in Fig.\ref{fig:5} 
the band structure of Ca$_4$Al$_2$O$_6$Fe$_2$As$_2$ 
for the hypothetical lattice structures 
varying the pnictogen height while fixing 
the lattice parameter $a$. Here the pnictogen height 
of 1.3\AA \ corresponds to $\alpha=110^\circ$ (close to the lattice 
structure of $x=1$) and 1.5\AA \ 
to 102$^\circ$ (close to $x=0$). 
In addition to the change of the Fermi surface 
configuration due to the the bond angle variation, the height reduction 
results in an increase of the band width (suppression of the density of 
states)  due to the reduction of the bond length\cite{Usui2011}.

\section{Superconductivity}
\subsection{FLEX approximation}

We now move on to the analysis of the spin fluctuation and 
superconductivity of Ca$_4$Al$_2$O$_6$Fe$_2$As$_2$. 
In addition to the tight binding model constructed from 
first principles band calculation, we consider the standard multiorbital 
interactions, namely, the intraorbital $U$, the interorbital $U'$, the Hund's 
coupling $J$, and the pair hopping interaction $J'$,  so the Hamiltonian reads,
\begin{eqnarray}
H &=& H_0 + \sum_i\left( \sum_\mu U_\mu n_{i\mu\uparrow} n_{i\mu\downarrow}
+\sum_{\mu > \nu}\sum_{\sigma,\sigma'}U'_{\mu\nu}n_{i\mu\sigma} n_{i\nu\sigma'}
\right.\nonumber\\
&&\left.-\sum_{\mu\neq\nu}J_{\mu\nu}\Vec{S}_{i\mu}\cdot\Vec{S}_{i\nu}
+\sum_{\mu\neq\nu}J'_{\mu\nu}
c_{i\mu\uparrow}^\dagger c_{i\mu\downarrow}^\dagger
c_{i\nu\downarrow}c_{i\nu\uparrow}
\right).
\end{eqnarray}
We apply the fluctuation exchange (FLEX) approximation\cite{Bickers1989,Dahm} 
using multiorbital 
Hubbard Hamiltonian. In FLEX, bubble and ladder type diagrams consisting of 
renormalized Green's functions are summed up to obtain the susceptibilities, 
which are used to calculate the self energy. The renormalized Green's 
functions are then determined self-consistently from the Dyson's equation.
The obtained Green's function is plugged into the linearized Eliashberg 
equation, whose eigenvalue $\lambda$ reaches unity at the superconducting 
transition temperature $T=T_c$.  Also, in order to investigate the correlation 
between superconductivity and magnetism, we obtain the Stoner factor $a_S$
of the antiferromagnetism at the wave vector $(\pi,0)$ in the unfolded 
Brillouin zone, which is defined as the largest eigenvalue of the matrix 
$U\chi_0({\bf{k}}=(\pi,0),i\omega_n=0)$, where $U$ is the interaction and 
$\chi_0$ is the irreducible susceptibility matrices, respectively. 
This value monitors the tendency towards stripe type antiferromagnetism and 
the strength of the spin fluctuations at zero energy. Since the three 
dimensionality is not strong in Ca$_4$Al$_2$O$_6$Fe$_2$As$_2$, 
we take a two dimensional model where 
we neglect the out-of-plane hopping integrals, and take $32 \times 32$ 
$k$-point meshes and  4096 Matsubara frequencies.

As for the electron-electron interaction values, 
we adopt the orbital-dependent interactions
as obtained from first principles calculation 
in ref.\onlinecite{Miyake_private} 
for Ca$_4$Al$_2$O$_6$Fe$_2$As$_2$, but multiply all of them 
by a constant reducing 
factor $f$. The reason for introducing this factor is as follows.
As has been studied in refs.\onlinecite{Ikeda_prb,Arita,Ikeda_jpsj} 
the FLEX calculation for models obtained from LDA calculations tends to 
overestimate the effect of the 
self-energy because LDA already partially takes into account the 
effect of the self-energy in the exchange-correlation functional. 
When the electron-electron interactions as large as those evaluated 
from first principles are adopted in the FLEX calculation, 
this double counting of the self-energy becomes so large 
that the band structure largely differs from its original one. 
In such a case, the spin fluctuations will develop around the wave vector
$(\pi,\pi)$ rather than $(\pi,0)$, which is in disagreement 
with the experimental observations. 
In the present study, we therefore 
introduce the factor $f$ so as to reduce the electron-electron interactions,
while maintaining the relative magnitude between interactions of different 
orbitals.

\subsection{Bond angle}

\begin{figure}
\includegraphics[width=8.5cm]{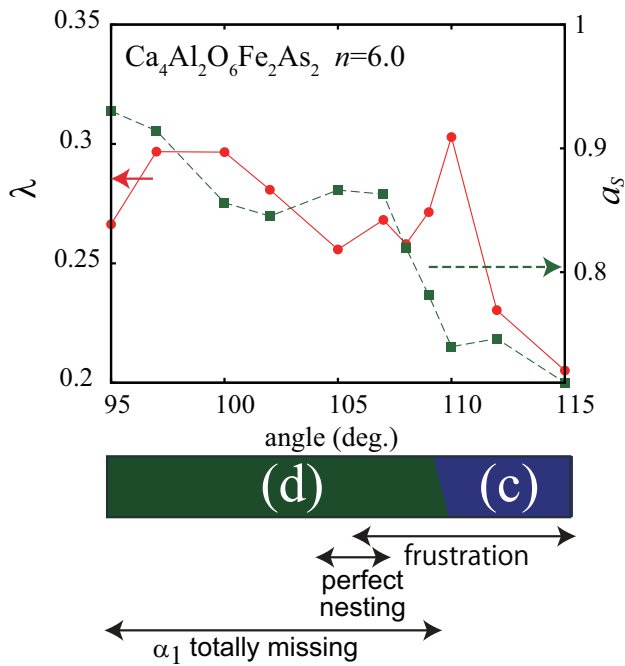}
\caption{(a) The Eliashberg equation eigenvalue for 
superconductivity ($s\pm$-wave pairing) (solid) and the Stoner factor 
at $(\pi,0)$ (dashed) against the bond angle for 
temperature $T = 0.005$. The interaction 
reduction factor is $f = 0.45$.\label{fig:6}}
\end{figure}

We show the eigenvalue of the Eliashberg equation $\lambda$ for 
the s$\pm$-wave superconductivity and the Stoner factor at $(\pi,0)$ for 
the hypothetical lattice structure of Ca$_4$Al$_2$O$_6$Fe$_2$As$_2$ varying 
the bond angle while fixing the bond length(Fig.\ref{fig:6}).

\begin{figure}
\includegraphics[width=8.5cm]{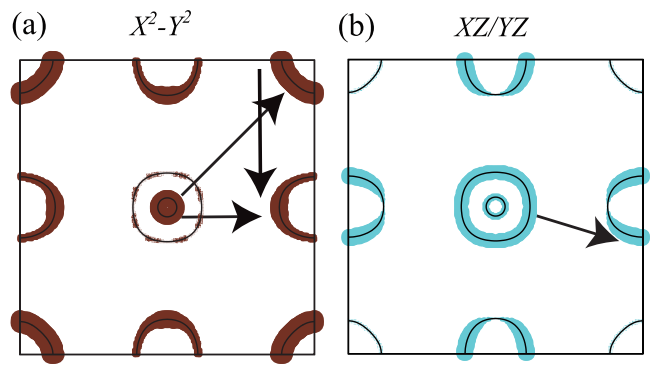}
\caption{The arrows indicate the wave vector of the 
dominant pairing interactions for the (a)$X^2-Y^2$ and 
(b) $XZ/YZ$ portions of the Fermi surface in the case where 
the inner hole Fermi surface ($\alpha_1$) is barely present. 
In this case, $\alpha_1$ is a mixture of $X^2-Y^2$ and $XZ/YZ$.\label{fig:7}}
\end{figure}

\begin{figure}
\includegraphics[width=8.5cm]{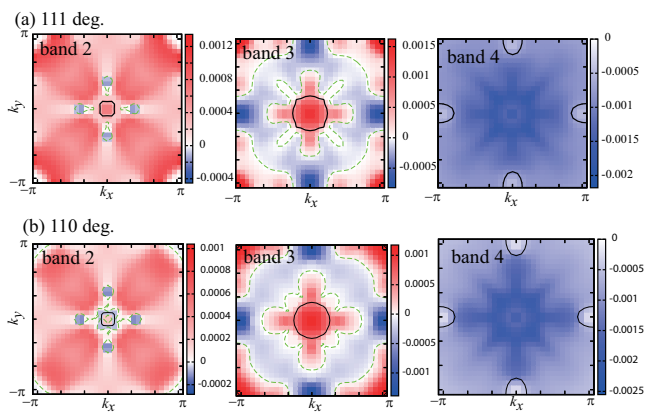}
\caption{The gap function obtained by FLEX for the hypothetical 
lattice structures of Ca$_4$Al$_2$O$_6$Fe$_2$As$_2$. The bond angle 
$\alpha$ is set to 110$^\circ$ or 111$^\circ$, while the bond length 
is fixed at the original value.\label{fig:8}}
\end{figure}

\begin{figure}
\includegraphics[width=8.5cm]{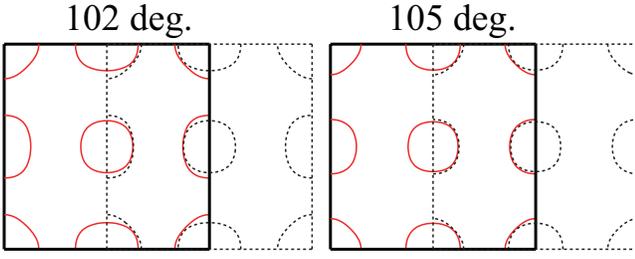}
\caption{The Fermi surface of Ca$_4$Al$_2$O$_6$Fe$_2$As$_2$  
for the hypothetical lattice structures with $\alpha=105^\circ$ 
and $102^\circ$ (solid), superposed with the Fermi surface 
shifted by $(\pi,0)$ (dashed).\label{fig:9}}
\end{figure}

As we decrease the bond angle from 115 to 110$^\circ$, 
eigenvalue of the Eliashberg equation $\lambda$ increases, reflecting 
the appearance of the $\gamma$ Fermi surface around $(\pi,\pi)$.
Superconductivity is locally optimized around 110$^\circ$,  
but $\lambda$ immediately goes down for larger bond angle. 
This is in contrast to the case of LaFeAsO, 
where $\lambda$ is broadly maximized around the regular tetrahedron 
bond angle.
This difference can be understood from the comparison between 
Fig.\ref{fig:2}(c) and Fig.\ref{fig:4}(b).
Namely, in the case of Ca$_4$Al$_2$O$_6$Fe$_2$As$_2$ with hypothetical 
bond angle, the Fermi surface configuration (b) 
with the optimal Fermi surface configuration 
is missing, i.e., in the three Fermi surface regime,  
$\alpha_1$ Fermi surface around $(0,0)$ is 
constructed from a mixture of $X^2-Y^2$ and $XZ/YZ$ orbital  
characters. In this configuration, The pair scattering takes place not only 
at $\sim (\pi,0)$ but also at $\sim (\pi,\pi)$ 
due to the same orbital character between 
$\alpha_2$ and $\gamma$ Fermi surfaces.
Since these Fermi surfaces interact with repulsive pairing interactions, 
a frustration arises in the sign of the superconducting gap 
as shown schematically in Fig.\ref{fig:7}. 
In addition to this, there can also be some $XZ/YZ$ component remaining in the 
$\alpha_1$ Fermi surface, and this portion tends to change the sign from
the $\beta$ Fermi surfaces, making it another possible factor
for the frustration.
The effect of the frustration appears in the form of the superconducting gap.
In Fig.\ref{fig:8}, we show the gap function for 
the hypothetical lattice structure of Ca$_4$Al$_2$O$_6$Fe$_2$As$_2$ 
at the bond angles 110$^\circ$ and 111$^\circ$. 
The sign of the gap function on $\alpha_1$ is positive at 111$^\circ$, 
but is very small (barely positive) at 110$^\circ$\cite{commentSUST},
reflecting the effect of the frustration.
The bond angle of 110$^\circ$ is actually very close to that of 
Ca$_4$Al$_2$O$_6$Fe$_2$P$_2$, so the appearance of a very small gap at 
this bond angle may be related to the nodal gap structure 
suggested experimentally for Ca$_4$Al$_2$O$_6$Fe$_2$P$_2$\cite{Kinouchi2011}.
As the bond angle is further reduced, the $\alpha_1$ Fermi surface 
disappears but the effect of the frustration remains strong as far as the 
top of the $\alpha_1$ hole band  does not sink far below the 
Fermi level. In fact, the frustration effect can be very strong 
right after the Fermi surface disappears because the top of 
this $\alpha_1$ band  (the closest point to the Fermi level) 
has pure $X^2-Y^2$ orbital character.
Therefore, $\lambda$ is suppressed around the bond angle of 
105$^\circ\sim$ 108$^\circ$. Meanwhile, the Fermi surface nesting itself
becomes very good in this regime because there are now two hole and two 
electron Fermi surfaces with no doped carriers, so that the 
average area of the hole and the electron Fermi surfaces becomes the same.
In particular, around the bond angle of 105$^\circ$, the nesting 
becomes nearly perfect, as shown in Fig.\ref{fig:9}. Therefore, 
the Stoner factor at $(\pi,0)$ takes a local maximum around this bond angle.
As the bond angle is reduced even further, the $X^2-Y^2$ band 
sinks far below the Fermi level and the frustration effect 
becomes small, so that $\lambda$ increases once again to a value 
comparable to that around the local maximum around the regular 
tetrahedron bond angle.
At the same time, the Fermi surface nesting becomes somewhat degraded, 
and the Stoner factor is reduced. 
For smaller bond angle$<96^\circ$ (which may not be realistic), 
the Fermi surface becomes too large, and the 
superconductivity is degraded. The bottom line here is that 
superconductivity is favored at around two bond angles 102$^\circ$ and 
110$^\circ$, and antiferromagnetism is favored in the regime in between 
these angles.
This is at least qualitatively consistent with the experimental 
observations for Ca$_4$Al$_2$O$_6$Fe$_2$As$_{1-x}$P$_x$.

The important point here is that 
superconductivity is suppressed in the intermediate bond angle regime 
due to the frustration effect. Apart from this, 
antiferromagnetism is favored around this bond angle 
regime due to a nearly perfect nesting of the Fermi surface.

\subsection{Pnictogen height}

\begin{figure}
\includegraphics[width=6.5cm]{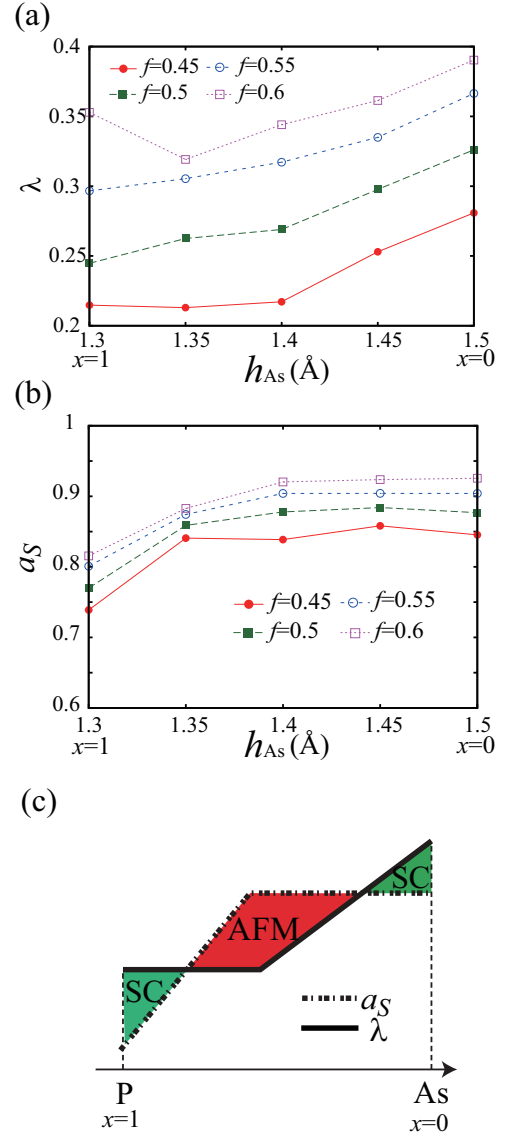}
\caption{The pnictogen height dependence of (a) the Eliashberg equation 
eigenvalue and (b) the Stoner factor at $(\pi,0)$ for the hypothetical 
lattice structure of  Ca$_4$Al$_2$O$_6$Fe$_2$As$_2$. 
Several values of the reducing factor are taken for comparison.\label{fig:10}
(c) A schematic figure of the $x$ dependence of $\lambda$ for 
superconductivity and $a_S$ for antiferromagnetism.}
\end{figure}

We have studied in the previous section the bond angle dependence of 
superconductivity and the spin fluctuations, and mentioned the 
possible relation between the calculation results and 
the experimental observations for Ca$_4$Al$_2$O$_6$Fe$_2$As$_{1-x}$P$_x$.
As mentioned previously, 
the actual lattice structure variation upon replacing As by P is 
more close to the variance of the 
pnictogen height $h_{\rm Pn}$ rather than just the bond angle. 
The increase of the bond length results in an increase in the density of 
states, generally resulting in an enhancement of 
both superconductivity and spin fluctuations\cite{Usui2011}.
In Fig.\ref{fig:10},  
we show the eigenvalue of the Eliashberg equation and 
the Stoner factor at $(\pi,0)$ for the hypothetical lattice 
structure of  Ca$_4$Al$_2$O$_6$Fe$_2$As$_2$ 
varying solely the pnictogen height $h_{\rm Pn}$.  
Around $h_{\rm Pn}=1.3\sim 1.35{\rm \AA}$, corresponding to the 
P content close to unity, 
the height dependence of $\lambda$ is weak (or $\lambda$ is even suppressed 
with the increase of $h_{\rm Pn}$ for large $f$),
while the Stoner factor rapidly increases with $h_{\rm Pn}$. 
This height regime corresponds to the bond angle regime of 
$110^\circ\sim 108^\circ$, where 
superconductivity is suppressed due to the momentum space 
frustration,  and at the same time antiferromagnetism 
is favored due to the nearly perfect nesting (Fig.\ref{fig:7}). 
Here in Fig.\ref{fig:10}(a), 
the enhancement of superconductivity 
by the increase of the density of states is canceled out due to the 
frustration effect, so that the $h_{\rm Pn}$ dependence of $\lambda$ is weak.
On the other hand, the Stoner factor quickly grows due to the 
cooperation of the 
good nesting and the increased density of states.
As the pnictogen height increases further beyond $1.35{\rm \AA}$, 
$\lambda$ starts to 
increase rapidly due to the reduction of the frustration 
and the increase of the 
density of states, while the Stoner factor tends to saturate because 
the nearly perfect nesting is degraded.
This overall tendency is summarized in a schematic figure in 
Fig.\ref{fig:10}(c)

\section{Pressure experiment}
Our theoretical study so far has 
shown that in the region where antiferromagnetism 
appears in the phase diagram, not only antiferromagnetism is 
enhanced due to the good Fermi surface nesting, but also superconductivity is 
suppressed due to the momentum space frustration, and these two are 
independent matters. Since superconductivity is suppressed 
regardless of whether antiferromagnetism is present or not,
superconductivity may not take place  
even when antiferromagnetism is suppressed by applying pressure, as 
is often done in other iron based superconductors. 

\begin{figure}
\includegraphics[width=8.5cm]{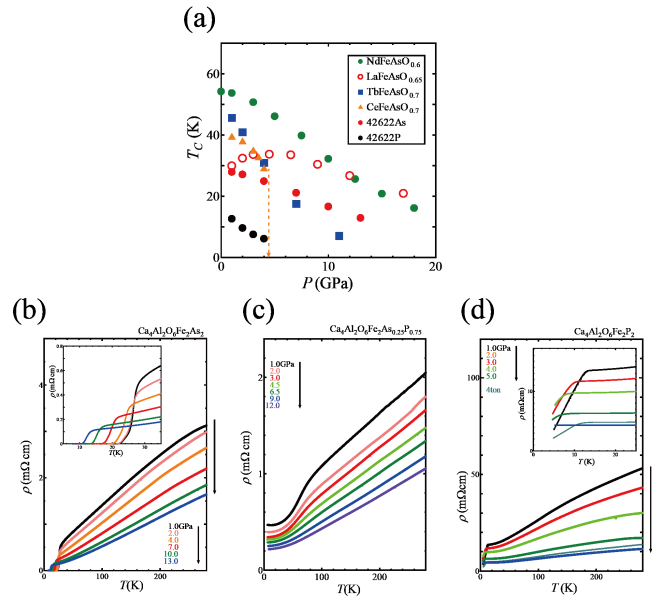}
\caption{(a) The pressure dependence of the superconducting 
transition temperature for various materials.  The resistivity 
against pressure for Ca$_4$Al$_2$O$_6$Fe$_2$(As$_{1-x}$P$_x$)$_2$ for 
(b) $x=0$, (c)$x=0.75$ and (d)$x=1$.  \label{fig:11}}
\end{figure}

To actually see this experimentally, 
we have applied hydrostatic pressure to 
Ca$_4$Al$_2$O$_6$Fe$_2$(As$_{1-x}$P$_x$)$_2$. 
The results are shown in Fig.\ref{fig:11}.
For the end compounds $x=0$ and $x=1$, $T_c$  
monotonically decreases with increasing pressure.
This is most likely due to the decrease in the density of states.
For $x=0.75$, where antiferromagnetism takes place at ambient pressure, 
superconductivity is not found up to 12GPa, although the 
antiferromagnetic transition is smeared out at high pressures. 
This is in contrast with cases where antiferromagnetism takes place at 
ambient pressure, but gives way to superconductivity under 
pressure.
The present experimental result supports the scenario that 
superconductivity in the intermediate $x$ regime is suppressed 
by momentum space frustration, apart from the presence of the 
antiferromagnetism itself.

\section{Conclusion}
In the present paper, we studied the origin of the peculiar phase 
diagram obtained for Ca$_4$Al$_2$O$_6$Fe$_2$As$_{1-x}$P$_x$ 
using a five orbital model constructed from first principles 
band calculation.
While the inner hole Fermi surface is absent at $x=0$\cite{Miyake2010}, 
it is present at $x=1$, but the orbital character has strong
 $X^2-Y^2$ character rather than $XZ/YZ$ as in LaFeAsO. 
This gives rise to momentum space frustration of the 
pairing interaction mediated by spin fluctuations, and degrades 
superconductivity. We propose this to be one of the reasons why 
$T_c$ is not so high in Ca$_4$Al$_2$O$_6$Fe$_2$P despite of the 
maximized multiplicity of the hole Fermi surface.
The frustration effect remains strong 
even after the inner Fermi surface has disappeared for $x<1$ 
because the top of the band with $X^2-Y^2$ orbital character 
remains near the Fermi level. At the same time, the 
disappearance of the most inner hole Fermi surface gives 
very good nesting of the electron and hole Fermi surfaces due to the 
equal number of sheets, favoring antiferromagnetism in the 
intermediate regime of $x$. Finally for $x\sim 1$, the top of the 
band sinks far below the Fermi level, and the frustration effect 
is reduced, so that superconductivity is favored once again.
Although we cannot directly determine which one of the 
superconductivity and antiferromagnetism wins, the tendency 
observed in the calculation is at least consistent with the 
experimental observation, where nodeless and nodal superconducting 
phases are separated by an antiferromagnetic phase.
Finally, we have performed hydrostatic pressure experiment, 
which further supports our scenario that superconductivity 
is suppressed by momentum space frustration in the intermediate 
$x$ regime.

\section{ACKNOWLEDGMENTS}

We are grateful to H. Mukuda, H. Kinouchi, and Y.Kitaoka 
for fruitful discussions.
The numerical calculations were performed at the Supercomputer Center, 
ISSP, University of Tokyo. This study has been supported by 
Grants-in-Aid for Scientific Research from JSPS. 
K.S acknowledges support from JSPS.

\bibliography{perovskite}

\begin{thebibliography}{99}
\bibitem{commentSUST} The result given here is slighly different from 
those presented in our previous paper\cite{SUST}. There, the band filling was 
taken at $n=6.1$, while in the present paper, the band filling is $n=6$.
\end{thebibliography}

\end{document}